\documentclass{cimento_mod}

\usepackage{graphicx} 

\title{Flaring Active Galactic Nuclei. The cases of 3C 279 and PMN
  J0948$+$0022 as seen by the Fermi-LAT}
\author{F.~D'Ammando\from{INAF-IASF Palermo} on behalf of the {\it
             Fermi}-LAT Collaboration}

\instlist{\inst{INAF-IASF Palermo} INAF-IASF Palermo, Via Ugo La
Malfa 153, I-90146 Palermo, Italy}
\PACSes{\PACSit{95.85.Pw}{Astronomical observations: $\gamma$ rays.}
\PACSit{98.54.Cm}{Active Galactic Nuclei}}

\begin{document}

\maketitle

\begin{abstract}
Active galactic nuclei (AGNs) exhibit variability across the entire electromagnetic spectrum with distinct flaring episodes at different frequencies.
The high sensitivity and nearly uniform sky coverage of the Large Area
Telescope on board the {\it Fermi}\, satellite make it a powerful tool for
monitoring a large number of AGNs over long timescales. This allowed us to
detect several flaring AGNs in $\gamma$ rays, triggering dedicated
multifrequency campaigns on these sources from radio to TeV energies. We
discuss the results for two different types of flaring AGN: the flat spectrum radio quasar 3C
279, in particular the coincidence of a $\gamma$-ray flare from this source with the drastic
change of the optical polarization angle, and the first
$\gamma$-ray flare from a radio-loud narrow-line Seyfert 1, PMN J0948$+$0022.   
\end{abstract}

\section{Introduction}

Since its launch on 2008 June 11, the {\it Fermi} Gamma-ray Space Telescope
has opened a new era in high-energy astrophysics. The primary instrument
on board {\it Fermi}, the Large Area Telescope (LAT), is a pair-conversion
telescope covering the energy range $\sim$ 20 MeV to 300 GeV with
unprecedented sensitivity and effective area~\cite{Atwood}. The combination of deep and
fairly uniform exposure over two orbits, very good angular resolution, and stable response of
the LAT has allowed it to produce the most sensitive, best-resolved survey of the
$\gamma$-ray sky, and to efficiently find episodes of flaring from $\gamma$-ray sources of different nature. 

One of the major scientific goals of the {\it Fermi} mission is to investigate
the high-energy emission in Active Galactic Nuclei (AGNs) in order to understand
the mechanisms by which the particles are accelerated and the precise site of
the $\gamma$-ray emission, and investigate on long timescales the AGN
variability and the $\gamma$-ray duty cycle. 
With respect to previous $\gamma$-ray instrument, such as EGRET~\cite{Thompson} and AGILE~\cite{Tavani}, the LAT
provides opportunities to investigate more in detail the behaviour of flaring $\gamma$-ray AGNs. Best examples are the extraordinary outbursts of
3C 454.3 in December 2009 and April 2010 (see~\cite{Escande} for details). Together with simultaneous multiwavelength observations
collected over the entire electromagnetic spectrum, LAT measurements allow us to reach a deeper insight
on the jet structure and the emission mechanisms at work in AGNs. In the
following we focus on two different types of AGN: the Flat Spectrum Radio Quasar (FSRQ) 3C 279, and the radio-loud Narrow-Line Seyfert 1 (RL-NLS1) PMN J0948$+$0022.

\section{The $\gamma$-ray flare/optical polarization change correlation in 3C\,279}

The FSRQ 3C 279 was one of the brightest blazar detected by EGRET with strong and rapidly variable $\gamma$-ray
activity~\cite{har01}, and the first FSRQ detected at energies above
100 GeV by the MAGIC Cherenkov telescope on February 2006~\cite{3C279MAGIC}. 
After a quiescent phase for the first $\sim$100 days of {\it Fermi}-LAT
operations, 3C\,279 entered a phase of strong $\gamma$-ray activity and a
multiwavelength campaign was triggered involving a large number of
observatories from radio to $\gamma$-ray bands, including also optical polarimetric observations by the Kanata telescope, as reported in detail in~\cite{Abdo2010a}. 

\begin{figure}[!thhh]
     \centering
   	\includegraphics[clip,width=0.65\linewidth]{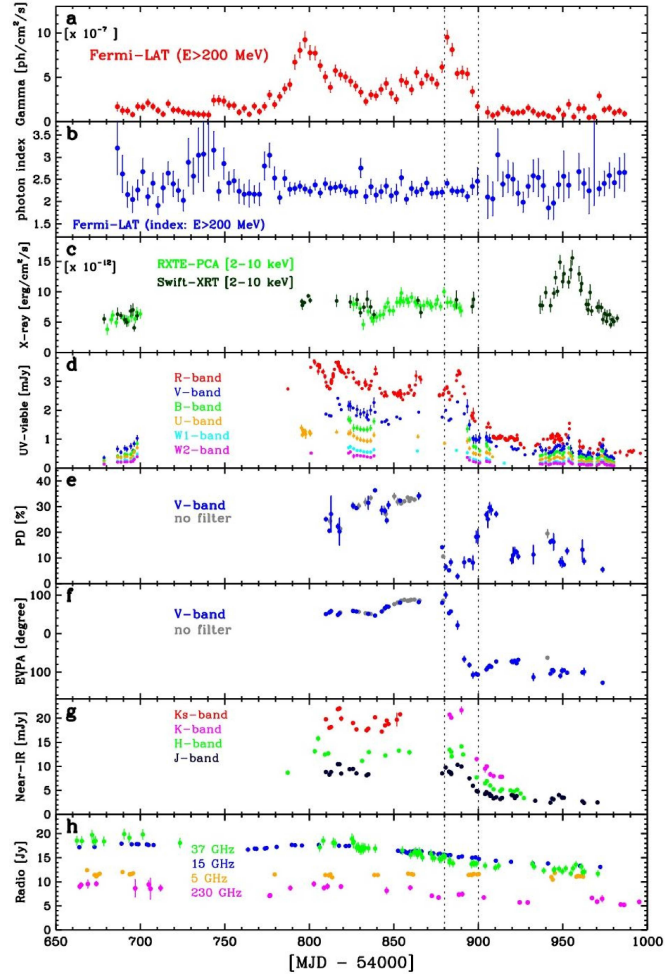}
	\caption{
Multifrequency light curves of 3C\,279 obtained between July 2008 and June
2009. {\it (a,b)}\,:~$\gamma$-ray flux above 200 MeV and photon index averaged
over 3-day intervals as measured by {\it Fermi}-LAT. {\it (c)}\,:~X-ray flux
integrated between 2 and 10 keV provided by {\it Swift}/XRT (dark green) and RXTE/PCA
(light green). {\it (d)}\,:~Optical and UV fluxes collected by GASP-WEBT, Kanata
and {\it Swift}/UVOT. {\it (e,f)}\,:~Polarization degree (PD) and electric vector
position angle (EVPA) of the optical polarization measured by Kanata in the V-band (dark blue) and by KVA Telescope with no filter (light blue). {\it
(g)}\,:~Near-infrared flux measured by Kanata and GASP-WEBT. {\it (h)}\,:~Radio
fluxes measured by OVRO at 15 GHz, and GASP-WEBT at 5, 37, and 230 GHz (Adapted
from~\cite{Abdo2010a}).} \label{3c279}
    \end{figure}

As shown in Fig.~\ref{3c279}, 3C\,279 went into a high $\gamma$-ray
state at around MJD 54780 lasting for about 120 days, a period characterized
by erratic flaring events and an overall double-peak structure with variations
of the flux by a factor of about 10. During the multiwavelength campaign the $\gamma$-ray
emission dominated the electromagnetic output of 3C 279 with an observed
$\gamma$-ray luminosity as much as $\sim$10$^{48}$ erg s$^{-1}$. 
%______________________________________________ 
The most striking event occurred during the rapid, second $\gamma$-ray 
flare at MJD 54880, with a doubling time scale of about one day. A highly correlated behavior of the $\gamma$-ray and optical bands is
evident between MJD 54880 and 54900, with the sharp $\gamma$-ray flare
coincident with a significant drop of the level of optical
polarization degree (PD), from about 30\% down to a few percent in $\sim$20
days. It suggests highly ordered magnetic fields in the $\gamma$-ray emission
region. This event is associated with a dramatic change
of the electric vector position angle (EVPA) that decreases by 208$^{\circ}$
with a rate of 12$^{\circ}$ per day. This behaviour
is in contrast to the relatively constant value observed before of about
50$^{\circ}$, corresponding to the jet direction of 3C\,279 as observed by
VLBI (see~e.g.~\cite{jor05}). This clearly indicates that the sharp
$\gamma$-ray flare is closely correlated with the dramatic change of optical
polarization due to a single, coherent event and the optical and $\gamma$-ray
emission regions are co-spatial.

The X-ray observations show a relatively constant flux during the second
$\gamma$-ray flare but reveal a symmetrical X-ray flare at about MJD 54950,
$\sim$60\,days after the $\gamma$-ray peak. During the X-ray flare optical and $\gamma$-ray emission is in a lower state, even if the
$\gamma$-ray emission is still dominant. That the X-ray spectrum is much harder than
the optical spectrum argues against a temporary extension of the high-energy tail of
the synchrotron emission. Moreover, the similarity of the X-ray shape and
time-scale with the $\gamma$-ray flare seems to be unfavourable to the
hypothesis that the X-ray flare is just a delayed version of the
$\gamma$-ray one due to, e.g., particle cooling. Thus, these findings are in favour of an isolated X-ray flare
produced by another mechanism that involves primary lower energy electrons,
challenging the simple one-zone emission models. Compared to the higher energy emission, the radio and millimeter fluxes are
less variable and no correlated or delayed radio flare was observed,
suggesting that the X-ray and $\gamma$-ray flaring events take places where
the radio emission is not yet fully optically thin. 
      
The gradual rotation of the optical polarization angle requires a
non-axisymmetric trajectory of the emission pattern or a swing of the jet
across our line of sight, since in the case of a uniform axially-symmetric
jet any compression of the plasma due to a perpendicular shock moving along the jet would result in a
change of the PD, but not in a gradual change of the EVPA, as instead we
observe. Two models have been proposed to explain the observed behavior in a
non-axisymmetric/curved geometry: the propagation of an emission knot along
helical magnetic field lines (see e.g.~\cite{Marscher2008}) or along the
curved trajectory of a ``bent jet'' (see e.g.~\cite{Young2010}). In both scenarios the distance of the
dissipation region from the central engine can be constrained from the $\sim$\,20\,day duration of the
polarization event to $\sim$ 10$^{5}$ gravitational radii away from the SMBH. This implies a jet opening angle of $<$\,0.2$^{\circ}$, smaller than 
typically observed with VLBI. This constraint could be relaxed in the
``flow-through'' scenarios, where the emission patterns may move slower than
the bulk speed of the jet or not propagate at all, and thus the modulation is due to the swing of the
whole jet across the line of sight. In this case the emission region could be
located at $\sim$ 10$^{3}$
gravitational radii, with the jet motion imposed at its base, caused by
deflection due to external medium, or as consequence
of dynamical instability. 

The dominant source of seed photons for the inverse-Compton contribution in
$\gamma$ rays depends on the distance of the dissipation region (broad line
region or accretion disk at sub-parsec scales, IR torus and synchrotron at
parsec scales) so it is fundamental to discriminate between the different
theoretical scenarios for understanding when the plasma blob dissipates in the jet. Long-term multiwavelength observations including {\it Fermi}-LAT and optical
polarization measurements of 3C 279 and other sources will be fundamental for providing new insights into the structure of relativistic jets of the blazars.

\section{PMN\,J0948+0022 and the $\gamma$-ray radio-loud Narrow-Line Seyfert
1s}

Relativistic jets are certainly the most extreme expression of the power than
can be generated by a SMBH in the center of an AGN, with bolometric luminosity up to
10$^{49-50}$ erg s$^{-1}$ (e.g.~\cite{Ackerman}), and a large fraction of the power emitted in $\gamma$ rays. Before the launch of the {\it Fermi} satellite, only two classes of AGNs
are known to generate these structures and thus to emit up to the $\gamma$-ray
energy range: blazars and radio galaxies, both hosted in giant elliptical
galaxies~\cite{Blandford}. The first 11 months of observation by
{\it Fermi}-LAT confirmed that the extragalactic $\gamma$-ray sky is dominated by these two classes~\cite{LAT11month}, but the discovery of
variable $\gamma$-ray emission from 4 RL-NLS1s revealed the presence of an
emerging third class of AGNs with relativistic jets~\cite{J0948_discovery,J0948_MW,4_NLS1}. This finding poses intriguing
questions about the knowledge of the development of relativistic jets, the
origin of the radio loudness, and the Unification model for AGNs.

NLS1 is a class of AGN discovered by~\cite{oster} and
identified by their optical properties: narrow permitted lines (FWHM
(H$\beta$) $<$ 2000 km s$^{-1}$) emitted from the broad line region, [OIII]/H$\beta$ $<$ 3, a bump due to FeII (see e.g.~\cite{pogge} for a review). They also exhibit strong X-ray variability, steep X-ray
spectra, and substantial soft X-ray excess. These characteristics point to systems with small masses of the central black hole (10$^6$-10$^8$ M$_\odot$) and high accretion rates
(up to 90$\%$ of the Eddington value) with respect to blazars and radio
galaxies. NLS1s are generally radio-quiet, with only a small fraction of them
($<$ 7$\%$,~\cite{komossa}) radio-loud, while generally 10$\%$-20$\%$
of quasars are radio-loud. In the past, several authors investigated the
peculiarities of RL-NLS1s with non-simultaneous radio to X-ray data,
suggesting similarities with young stage of quasars or different types of blazar \cite{komossa, yuan, Foschini09a}. The strong and variable radio emission, and the flat
radio spectrum together with variability studies suggested the presence in
some RL-NLS1s of a relativistic jet, now confirmed by the detection by
{\it Fermi}-LAT of $\gamma$-ray emission from PMN J0948$+$0022~\cite{J0948_discovery,Foschini09b}. After that source other 3 RL-NLS1s were
detected by LAT in $\gamma$ rays: 1H 0323$+$342, PKS 1502$+$036, and PKS 2004$-$447. The raising number of
$\gamma$-ray detection suggests that they form a new class of $\gamma$-ray emitting AGNs~\cite{4_NLS1}.  
Moreover, as in the case of blazars and radio galaxies, there should be
 a ``parent population'' with the jet viewed at large angles. The first
 source of this type was recently detected: PKS 0558$-$504~\cite{gliozzi}. Therefore, it seems that NLS1s could be a set of
 low-mass systems parallel to blazars and radio galaxies. Taking into account
 also the high accretion rate, the RL-NLS1 is a class of extreme interest for
 extending the studies of the properties of relativistic jets to different mass and power scales.

%-------------------------------------------------------------
   \begin{figure}[t]
   \centering
   \includegraphics[clip,width=0.60\linewidth]{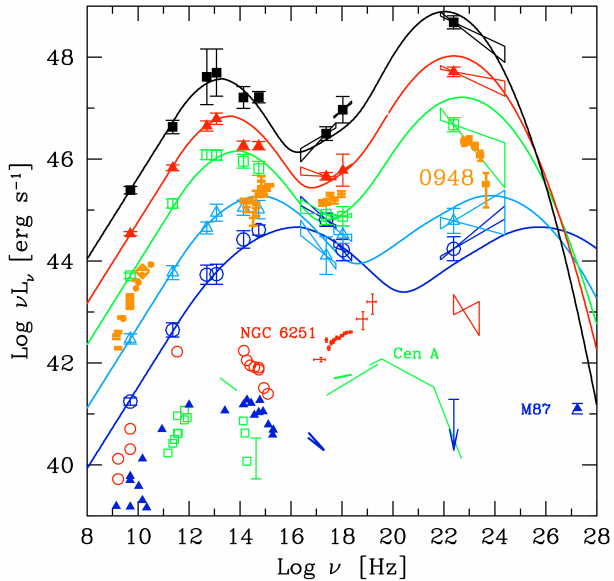}
\hfill\parbox[b]{0.38\textwidth}{\caption[]{
%      \caption{
The average SED of PMN\,J0948+0022 as obtained during the first 5 months of {\it
      	       Fermi} observation together with {\it Swift} (XRT and UVOT), OVRO and
      	       Effelsberg data (orange squares), here shown in comparison with
      	       the blazar sequence and the SEDs of the radio galaxies NGC
      	       6251, Cen A, and M87 (adapted from~\cite{Foschini09b}, see also~\cite{J0948_discovery,Ghisellini2005}).}} \label{0948_SED}
   \end{figure}
%------------------------------------------------------------- 

The $\gamma$-ray detection of these RL-NLS1s was important not only for the
 confirmation of the presence of a relativistic jet, but also to measure its
 power and study the characteristics of this class of objects by modeling
 the broad band spectrum. The first SEDs collected for these 4 RL-NLS1s showed clear similarities with blazars: a double-humped shape with
 disk component in UV (except PKS 2004$-$447), physical parameters blazar-like and jet power in the
 average range of blazars. For example, the physical parameters resulting from the modeling of the average SED of PMN J0948$+$0022 during the
multifrequency campaign in August 2008-January 2009 (see~\cite{J0948_discovery}) are typical of a source midway between FSRQs and
BL Lacs, with low power with respect to FSRQs (see Fig.~\ref{0948_SED}). A similar behaviour was observed also for the other three
objects. In particular, compared with the blazars, the jet power estimated for
 PMN 0948$+$0022 and PKS 1502$+$036 are in the region of FSRQs, while 1H
 0323$+$342 and PKS 2004$-$447 are in the range typical of BL Lac objects (see~\cite{4_NLS1}). This discovery challenges the blazar sequence and more generally the AGN Unification model. 
  
 Another key question was the maximum power released by the jets of RL-NLS1s. A first answer arrived in July 2010 when PMN J0948+0022 underwent two
 strong $\gamma$-ray flares with peak flux of $\sim$100 $\times$ 10$^{-8}$
 ph cm$^{-2}$ s$^{-1}$~\cite{Donato,Foschini10a}, corresponding to
 an apparent isotropic $\gamma$-ray luminosity of $\sim$10$^{48}$ erg s$^{-1}$, comparable to that of the
 bright FSRQs~\cite{Foschini10b,Foschini10c}. The extreme power of these flaring episodes
 confirms that this RL-NLS1 hosts relativistic jets as powerful as those in
 blazars, despite the relatively low mass and the rich environment due to the
 high accretion rate. The continuous monitoring of the entire $\gamma$-ray
 sky provided by {\it Fermi}-LAT allows us to catch, if it happens, a new
 intense $\gamma$-ray flare from one of these sources. It is
 important to know if also other RL-NLS1s can produce similar $\gamma$-ray
 flares to determine if PMN J0948$+$0022 is an archetypical source of this new
 class of $\gamma$-ray AGNs or if it shows peculiar characteristics also with
 respect to the other RL-NLS1s. Furthermore, by considering that NLS1s are
 usually hosted in spiral galaxies the presence of a relativistic jet in these objects is contrary to the paradigm that the
 formation of relativistic jets could happen only in elliptical galaxies (see
 e.g.~\cite{Bott,Marscher2009}). This suggests that relativistic jets can form and
 develop independently of their host galaxies, challenging our actual knowledge on
 the relativistic jet formation.  
   
\section{Conclusions}
The $\gamma$-ray all-sky monitoring by {\it Fermi}-LAT, combined with other
simultaneous ground- and space-based observations, provide us new insights into
the relativistic jets and broad emission models of AGNs. We presented the
results from the multifrequency campaign of the FSRQ 3C 279, including the
discovery of a $\gamma$-ray flare associated with a drastic change of the
optical polarization angle as well as the detection of an ``orphan'' X-ray
flare. In addition, the discovery by {\it Fermi}-LAT of $\gamma$-ray emission
from 4 RL-NLS1s provided evidence for relativistic jets in these
systems. The $\gamma$-ray flare of PMN J0948$+$0022 in July 2010 confirmed that extreme power can be produced also in this class of AGNs.

\acknowledgments

The {\it Fermi} LAT Collaboration acknowledges support from a number of
agencies and institutes for both development and the operation of the LAT as
well as scientific data analysis. These include NASA and DOE in the United
States, CEA/Irfu and IN2P3/CNRS in France, ASI and INFN in Italy, MEXT, KEK,
and JAXA in Japan, and the K.~A. Wallenberg Foundation, the Swedish Research
Council and the National Space Board in Sweden. Additional support from INAF
in Italy and CNES in France for science analysis during the operations phase is also gratefully acknowledged.

\end{document}